\documentclass{article}
\usepackage[a4paper,top=3cm,bottom=2cm,left=3.5cm,right=3.5cm,marginparwidth=1.75cm]{geometry}
\usepackage{mathpazo}
\pdfoutput=1
\usepackage{float}

\usepackage{ragged2e}
\usepackage{bm}
\usepackage{amsmath}
\usepackage{tabularx,ragged2e,array}

\usepackage{booktabs}

\usepackage[utf8]{inputenc}
\usepackage[T1]{fontenc}
\usepackage{placeins}
\usepackage{siunitx} % centering in tables
\def\sym#1{\ifmmode^{#1}\else\(^{#1}\)\fi}

\usepackage{graphicx}
\usepackage[labelformat=empty]{caption}
\usepackage{booktabs}
\usepackage[gen]{eurosym}
\usepackage{gensymb}
\usepackage{hyperref}
\usepackage{lscape}
\usepackage{dirtytalk}

\def\sym#1{\ifmmode^{#1}\else\(^{#1}\)\fi}

\usepackage{hyperref}
\usepackage[hang,multiple]{footmisc}
\usepackage[T1]{fontenc}

    \usepackage[format=hang,font=bf,labelfont=bf]{caption}
\usepackage{authblk}
\usepackage{lineno}
\usepackage{dirtytalk}

\usepackage{tcolorbox}

\usepackage[
  style       = nature,
  autocite    = superscript,
  backend     = biber,
  sortcites   = true,
  sorting = none,
]{biblatex}
\title{\Large{Effect of pop-up bike lanes on cycling in European cities}}

\author[1,3]{\normalsize{Sebastian Kraus\thanks{Corresponding author. Email address: kraus@mcc-berlin.net. ORCID: https://orcid.org/0000-0003-1161-2988}}}
\author[1,2]{Nicolas Koch}
\affil[1]{{\footnotesize Mercator Research Center on Global Commons and Climate Change, Torgauer Str. 19, 10829 Berlin, Germany}}
\affil[2]{{\footnotesize Potsdam Institute for Climate Impact Research, Telegrafenberg, 14473 Potsdam, Germany}} 
\affil[3]{{\footnotesize Technical University of Berlin, Straße des 17. Juni 135, 10623 Berlin, Germany}} 

\begin{document}
\maketitle
\begin{refsection}

% \textbf{\Large{Summary}}

% \paragraph{Background} The bicycle is a low-cost means of transport linked to low risk of COVID-19 transmission. Governments have incentivised cycling by redistributing street space as part of their post-lockdown strategies.
% \paragraph{Methods} We evaluated the impact of provisional bicycle infrastructure on cycling traffic in European cities using a generalised difference-in-differences design. We scraped daily bicycle counts spanning over a decade from 736 bicycle counters in 106 European cities. We combined this with data on announced and completed pop-up bike lane road work projects.
% \paragraph{Findings} On average 11.5 kilometres of provisional pop-up bike lanes have been built per city. Each kilometre has increased cycling in a city by 0.6\%. We calculated that the new infrastructure will generate \$2.3 billion in health benefits per year, if cycling habits are sticky.
% \paragraph{Interpretation} Provisional cycling infrastructure can induce fast, substantial shifts to a sustainable mode of transport with low risk of COVID-19 transmission at a low cost.

% \paragraph{Funding} This study was not externally funded.
% \vspace{0.3cm}

\begin{abstract} The bicycle is a low-cost means of transport linked to low risk of COVID-19 transmission. Governments have incentivised cycling by redistributing street space as part of their post-lockdown strategies. We evaluated the impact of provisional bicycle infrastructure on cycling traffic in European cities using a generalised difference-in-differences design. We scraped daily bicycle counts spanning over a decade from 736 bicycle counters in 106 European cities. We combined this with data on announced and completed pop-up bike lane road work projects. On average 11.5 kilometres of provisional pop-up bike lanes have been built per city. Each kilometre has increased cycling in a city by 0.6\%. We calculate that the new infrastructure will generate \$2.3 billion in health benefits per year, if cycling habits are sticky.
\end{abstract}

\section{Introduction}
As social and economic activity resume after a period of social distancing to curb COVID-19, policy-makers are seeking mitigation measures with favourable cost-benefit ratios that can be implemented in the short-run. While overall mobility is almost back to pre-crisis levels in many European countries, the use of public transport is still lagging behind\autocite{apple2020}. Early evidence points to shifts from public transport to car use as users react to the pandemic\autocite{chang2020}. Governments have started incentivising cycling as a low-cost, sustainable, equitable, and space-saving mode of transport that reduces the risk of COVID-19 transmission. A key measure has been the redistribution of street space in cities to create provisional bike infrastructure often provisionally marked and protected by materials readily available from road construction companies. As of 8 July 2020, 2000 kilometres of these infrastructure changes had been announced\autocite{ecf2020}.

In Europe, typically more than 50\% of overall trips measured in transport surveys are shorter than 5 kilometres\autocite{ahern2013}. In 2019, 3 million electric bicycles were sold in the EU\autocite{leva2020} likely making cycling more demographically diverse and increasing the distances travelled\autocite{moser2018}. This speaks to an important short-term potential for shifts in transport mode choice that could reduce crowds in public transport and help avoid traffic congestion in response to increased car use out of fears of infection. 

Mode choices are subject to behavioural effects, such as status quo bias, default effects, and time-inconsistent preferences\autocite{mattauch2016}. This complicates the task of policy-makers to encourage people to cycle, particularly in the short-run. However, major disruptions to public transport, such as strikes, cause people to reconsider their habits\autocite{larcom2017}. Furthermore, highly visible, large-scale expansions in the provision of bicycle amenities, such as bike sharing\autocite{hamilton2018} or a city-wide network of 120 kilometres of separated bike lanes built within four years in Sevilla\autocite{marques2015}, have increased cycling and reduced congestion\autocite{hamilton2018}.

Here, we provide causal estimates of the effect of the post-COVID-19-lockdown roll-out of provisional (\say{pop-up}) bike lanes in European cities. We compile new data on daily bike counts in 110 cities. We connect to the open data application programming interfaces (APIs) of these cities to download bike counts from a total of 736 counters spanning over a decade. We combine this data with information on day-to-day changes in the number of kilometres of pop-up bike lanes, which is collected by the European Cyclists' Federation based on official documents and media reports.\autocite{ecf2020} Our sample consists of large and medium-sized cities in 20 European countries.

\section{Methods Summary}
We estimate a Poisson regression model at the counter level with daily counts of cyclists as the outcome variable and the number of kilometres of pop-up bike lanes in service in a city on a given day as the treatment.

Since the roll-out of pop-up bike lanes is not a controlled experiment, our main empirical concern is that both the implementation of bike lanes and bicycle counts are driven by a third factor that cannot be measured (omitted variable bias). We may also worry that bike lanes are built as a reaction to increased cycling traffic (reverse causality). We address these concerns using quasi-experimental variation in the roll-out of pop-up bike lanes in different European cities.

Planning for provisional cycling infrastructure in Europe has started early in the pandemic as a reaction to civil society pressure after announcements by the City of Bogota on 16 and 17 March to create 76 km of provisional bike lanes that was widely reported in the international media. Similar plans were assembled in several European cities and the roll-out of these plans has started during lock downs as a means to allow necessary travel under high safety standards particularly for \say{key workers} (see Fig. 1).

\begin{figure}[h!]
\centering
    \includegraphics[width=0.7\textwidth]{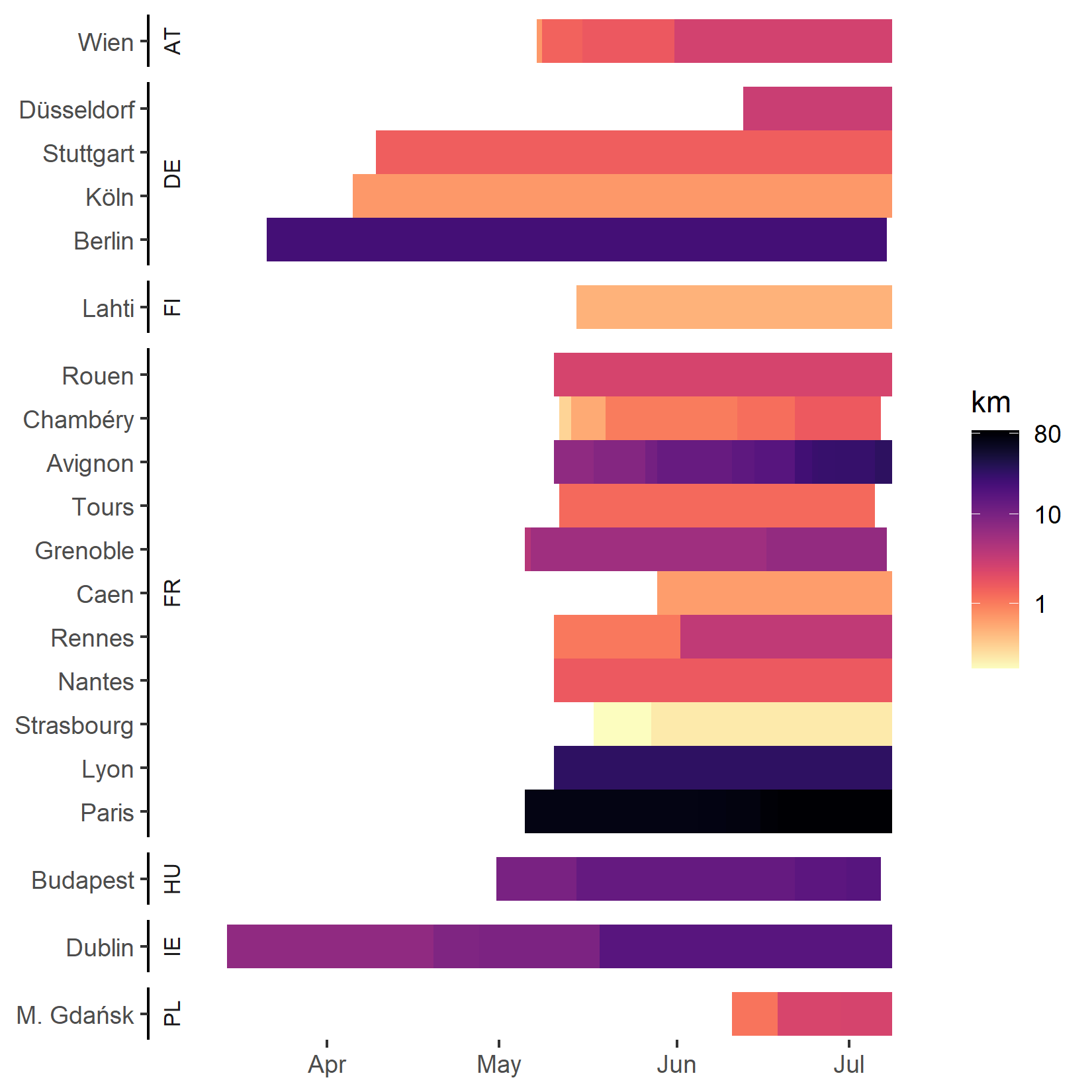}
    \caption{\scriptsize{{\bf Fig. 1. Intensity of pop-up bike lane treatment over time} This Figure shows treated cities and their treatment intensities in implemented kilometres of public bike lanes at a given day between March and July 2020. Control cities are not plotted but are included in Fig. S1 (see Supplementary Materials). London, Milan, Rome, and Lisbon are missing from the sample due to a lack of daily bicycle counter data. Bars that do not cover the whole study period to July 8 2020 indicate missing bicycle count data for the most recent dates due to updating time lags of the counter APIs.
    Information on individual pop-up bike lanes with their street location, announcement date, and implementation date is from the European Cyclists' Federation. The newest data can be found at: https://ecf.com/dashboard}}
\end{figure}

Officials have stated in interviews and personal conversations, that the geographic placement of pop-up bike lanes has mainly been driven by the availability of street space that can be redistributed without restricting car traffic to only one direction and the existence of \say{shovel-ready} construction plans. The exact timing of pop-up bike lane construction is driven by administrative idiosyncrasies and the availability and schedules of construction firms. Therefore, we argue that the timing of the roll-out of pop-up bike lanes has been as good as random.

Our regression analysis is based on comparisons between treatment and control groups before and after treatment around each cohort of new bike lanes (differences-in-differences). We use a set of indicator variables (fixed effects) that remove variation from our estimation sample that could be biasing our estimates. Our study design allows for systematic differences in the level of bike traffic between treatment and control group, but relies on a common trends assumption, that bike traffic in treated and control cities would have evolved on a parallel trend in the absence of treatment. Since we cannot observe treated units in their untreated state after treatment (potential outcome), we cannot test the common trends assumption formally. However, we can investigate pre-treatment trends and check the sensitivity of our estimates to changes in the control group definition.

As a baseline our difference-in-differences model includes fixed effects at the unit (counter) and time (day) level. We thereby control for time-invariant factors at the level of each counter and city, such as public transport and population density, topography, and preferences for green lifestyles. With our counter fixed effect we also rule out that our effect is driven by new counters that get placed next to provisional bike lanes. The day fixed effect removes trends from the treatment and outcome variation that are common to the whole sample. These could be overall trends in cycling, seasonality, and the overall evolution of the COVID-19 pandemic in Europe. Fixed effects at the country-day level remove variation in cycling infrastructure and behaviour that is driven by state- or national-level COVID-19 policies.

For potentially biasing factors that vary at the city-level over time, such as local mobility or weather, we cannot include fixed effects since this is the geographical level at which our treatment is measured. We therefore include control variables in our regressions that measure overall changes in mobility at the state-level. This variable is based on the aggregated movements of Facebook users. This is to rule out that our effect is driven by local authorities reacting to increased traffic volumes. We also control for local temperatures, sunshine, wind and precipitation. Weather could for instance create bias, when construction firms decide to create new bike lanes in weeks with good weather that will also have more cycling.

\section{Results}
We compare bike traffic in treated cities in the days before and after they get treated compared to control cities and find that one kilometre of popup bike lane increases cycling by 0.6\% (see right one of coefficients marked in blue in Fig. 2). When we multiply this estimate for a kilometre of bike lane with the average number of kilometres (11.5), we find that the average effect of bike lane programs is a 7\% increase in city-wide cycling.

\begin{figure}[h!]
\centering
    \includegraphics[width=0.5\textwidth]{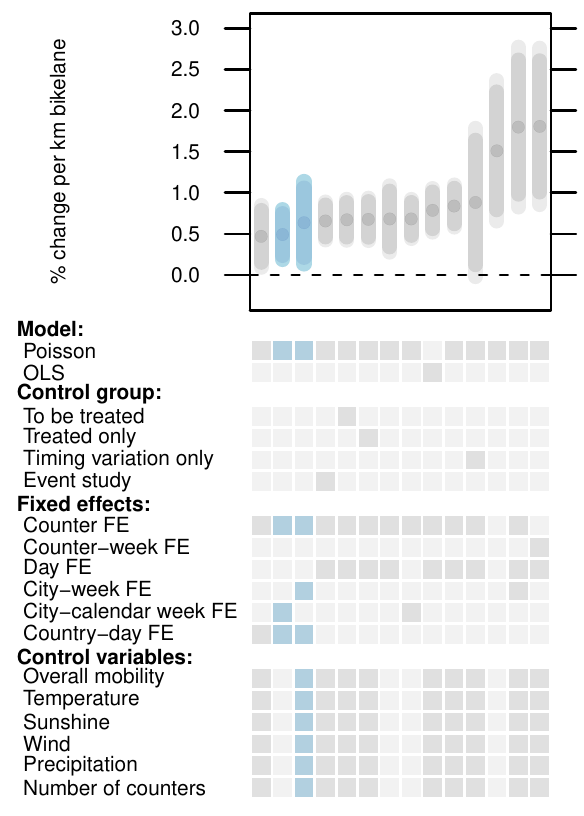}
    \caption{\scriptsize{{\bf Fig. 2. Effect of pop-up bike lanes on cycling in different model specifications} This figure shows estimates of regressions of the daily cyclist count on the number of kilometres of pop-up bike lane implemented at a given day in a city. The unit of observation is the bike counter. Baseline specifications are marked in blue. Darker colours in the bottom panel indicate the type of specification used. The 95\% confidence interval is shown in darker colour and the 90\% confidence interval in lighter colour. One estimate is from an OLS specification and uses the natural logarithm of the bicycle count as the outcome. All other specifications are Poisson regressions. The estimates can be interpreted as the average increase in the level of cycling caused by one kilometre of pop-up bike lane. Control variables are from Facebook (mobility index measured by user movements) and the ERA5 climate model (weather variables). The variable \textit{Number of counters} indicates the total of counters per city.}}
\end{figure}

The bike count data spans over a decade. We can compare changes in cycling in the weeks after the introduction of pop-up bike lanes with the same calendar weeks in previous years. Fig. 2 shows a regression estimate based on this comparison (see left one of coefficients highlighted in blue). Comparisons between weeks in 2020 and weeks in previous years may be biased by differential trends between treatment and control group. Fig. 3 shows the estimated difference between treatment and control group in the 12 months before and 3 months after the begin of the pop-up bike lane roll-out in March. The baseline category in this event study specification is $-13$. This means that all estimated coefficients for months before and after treatment are relative to February 2019.

\begin{figure}[h!]
\centering
    \includegraphics[width=0.7\textwidth]{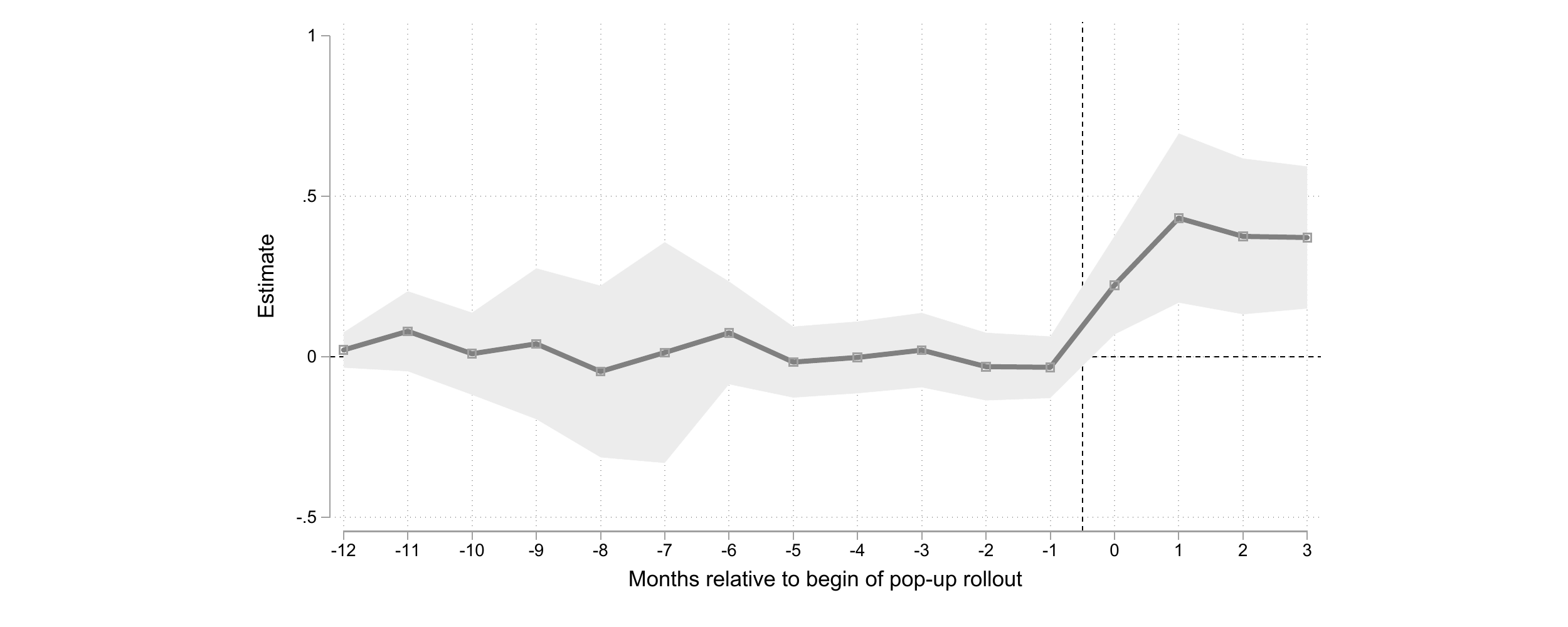}
    \caption{\scriptsize{{\bf Fig. 3. Treatment effect in months before and after beginning of pop-up bike lane treatment} This figure shows the treatment effect of treated cities compared to untreated cities. Observations are binned into months before and after treatment. The treatment is hard-coded to March 2020. The baseline category and the begin of the sample is February 2019. Estimates are from Poisson regressions that include city and country-day fixed effects. The shaded area shows the 95\% confidence interval.}}
\end{figure}

We can see that a treatment effect becomes only apparent after the treatment sets in. Before, treatment and control group have been on the same trend. There is a slight but statistically insignificant downward trend before treatment (Ashenfelter's dip\autocite{ashenfelter1978}), hinting at the possibility of stronger mobility reductions due to COVID-19 in cities that have decided to build pop-up bike lanes. This could be the case because local and national governments are more likely to take wide-ranging action, if their country is hit by a more intense outbreak. It could also be due to governments acting upon idiosyncratic risk-aversion of their populations towards cycling in the context of emptier roads and increased speeding during the lockdown. We rule out that these potential selection into treatment effects are driving our results by controlling for COVID-19 related dynamics with fixed effects and a variable that captures human mobility at the sub-national level based on Facebook user movements. 

The treatment effect magnitude in Fig. 3 is higher than our baseline estimate. This difference stems from hard-coding the treatment in March 2020 and therefore discarding variation in treatment timing at the day and week levels. This creates a more standard difference-in-difference setting, that avoids the issue of already treated cities acting as controls for later cohorts, while they are still on a different trend because of prior treatment\autocite{goodman-bacon2018}. Therefore our main estimates tend to be attenuated compared to the setup shown in Fig. 3.

We check the sensitivity of our results to reshaping our regression-based treatment and control group comparisons. Fig. 2 shows fairly stable estimates for comparisons between (i) treated and untreated cities, (ii) cities that are already treated and those that have only announced pop-up bike lanes, (iii) between treated cities only using their variation in treatment dose (km of bike lane built) and treatment timing or (iv) treatment timing only (event study).

Estimates based on days before and after treatment within the same week have higher magnitudes when country-day fixed effects are excluded from the model. This suggests that within a narrow time window around treatment, national policy events and cycling behaviour are correlated. Estimates that either include country-day fixed effects or use longer pre- and post-treatment windows (no city-week fixed effect or city-calendar week fixed effect only) mitigate this bias.

\section{Discussion}
We find robust evidence for substantial short-run increases in cycling in European cities due to new provisional cycling infrastructure. An average pop-up bike lane program has led to a 7\% increase in city-wide cycling. The effects of this cycling infrastructure on COVID-19 transmission should be investigated with high-resolution case data for a large enough number of cities. 

Independent of its impacts on COVID-19 transmission, the net benefits of the intervention are likely to be large. The direct cost of cycling infrastructure including planning is low. For the Sevilla network one kilometre of bike lane cost \euro{250000}\autocite{marques2015}. Iterative planning with provisional infrastructure reduces costs further. In Berlin, one kilometre of pop-up bike lane has so far cost \euro{9500}\autocite{bezirksamtfriedrichshain-kreuzberg2020}. Previous research has found that every kilometre of cycling generates health benefits of \$0.45\autocite{zapata-diomedi2018}. We calculate baseline values for total cycling in a city based on data on daily kilometres cycled in German cities in 2018 and extrapolate these numbers to the rest of our sample based on city-level data on modal splits and population (see Supplementary Material). We calculate that the additional cycling caused by the pop-up bike lane treatment during its first three months of operation has generated at least \${580} million in health benefits\footnote{Note, that our sample does not include infrastructure built after the July, 8th and excludes a small number of important cities, for which adequate open bike counter data is missing.}. The new infrastructure will generate \${2.3} billion per year in health benefits, if the new bike lanes become permanent and if cycling habits are sticky.

The magnitude of our estimate is large compared to previous evaluations of cycling infrastructure improvements that have found statistically unclear or modest effects, typically because of the limited scale of the interventions \autocite{yang2010,winters2017,aldred2019a}. Further research could investigate the non-linearities in cycling adoption in terms of scale and timing of an infrastructure roll-out. It remains to be evaluated, if cycling behaviour is sticky and how similar treatments influence behaviour outside of the pandemic environment.

Research based on surveys indicates that separated, protected infrastructure is a key element to incentivise up-take of cycling\autocite{manaugh2017,aldred2017}. Cities have experimented with a range of measures to create new spaces for cycling, ranging from painted to provisionally protected bike lanes and from traffic calming with signs to built \say{modal filters} that only let bicycles and pedestrians pass. In our data, we do not see which share of increased bike counts is from new cyclists and which is from existing cyclists, who decide to cycle more often or farther. Large representative individual level samples, for instance based on transport mode detection by smartphone sensors, may help to investigate changes in modal split at a sufficiently high geographical resolution. GPS traces of individual trips could also help understand, how new infrastructure changes the route choices of cyclists\autocite{hong2019} and to measure the willingness to take detours for better infrastructure in terms of value of time.

\paragraph{Contributions} S.K. ran the analyses. S.K. and N.K. designed the analysis, interpreted results, designed figures and wrote the paper. 
\paragraph{Declaration of interests} We declare no competing interests
\paragraph{Acknowledgements} We thank Ben Thies and Lennard Naumann for their excellent research assistance. We thank Jill Warren and Aleksander Buczyński at the European Cyclists' Federation (ECF) for their data, that can also be browsed at their dashboard\autocite{ecf2020}.
We thank numerous volunteers, that have contributed to this data collection effort. We thank Eco-Counter for their technology allowing cities to share their cycling counts publicly. We also thank Ariel Ortiz-Bobea for sharing his code to produce Specification Charts.

\paragraph{Data and materials availability:} Data and code used in our analysis are available at \autocite{kraus2020}.

\newpage
% \section{Figures}
\printbibliography
\end{refsection}

\clearpage

\newpage
\begin{refsection}
\section*{Supplementary Material}
\subsection*{Materials and Methods}
% \subsection*{Supplementary Information}
\paragraph{Bicycle count data}
We assemble a new data set of daily bicycle counts from municipal bicycle counters. We connect to national and municipal open data portals for bike counter data sets and connect directly to the API of those cities that use the Eco-Counter standard (see . We also obtain longer time series of bike counts going back to 2012 directly from the Mayor's staff for road planning and data in Paris. 

Our raw data set contains roughly a million daily counts starting in 2007. We drop the lower and upper percentiles from this raw sample since counters can record very low values, when they are not functioning properly or very high values, when there is a cycling event that drives up counts. We drop the counter 100041252 from Bergen that varies between very low values and some of the highest daily counts in the sample. Our results are robust to keeping these extreme values in the sample. The bulk of the bike counts are from most recent years (see Table S1) and we focus most of the comparisons made in our regressions on the years 2019 and 2020. Figures S3 and S4. show the variation in weekly average bike counts for cities in our study sample. Fig. S3. shows treated cities and Fig. S4. control cities. For certain cities, such as Paris and Berlin the raw data already indicates that increase in peak in June 2020 compared to June 2019. Many of the control cities show a similar pattern. Our regression analyses find a robust effect of new infrastructures, both when taking the difference in these differences between treatment and control cities, but also when focusing on variation in treatment timing exclusively.

Table S1 shows summary statistics for the main variables included in our analysis. The unit of observation in our analyses is the bike counter and counts vary daily. An average counter detects 1457 cyclists per day. The average number of counters per city is 22.9. The average size of cities in our sample is 33000 ha. European cities tend to be denser than American cities. Thus, our study areas can be thought of small commuting zones rather than city cores.

\paragraph{Pop-up infrastructure data}
We use project-level data on provisional infrastructure in European cities as a reaction to the COVID-19 pandemic collected by the European Cyclists' Federation\autocite{ecf2020}. In the data we see the street, where the project is implemented, its size measured in kilometres, the date of announcement, and the date of implementation. The data also contains the type of project. 80\% are categorised as bike lanes and 16\% as traffic calming. Our data includes all projects recorded until 8 July 2020. We aggregate this data at the city-day level to construct a variable of daily implemented kilometres of pop-up bike lane. We use the city definition and corresponding polygons from the European Urban Audit 2020\autocite{eurostat2020a}. Typically areas defined by the European Urban Audit include suburbs. For instance, the Paris polygon includes many areas beyond the ring highway that surrounds the municipality of Paris ("Ville de Paris"). This allows us to capture commuting enabled by new bike lanes from the suburbs into the city centre, which make up an important share of projects (see Fig. S1). However, this also means that for infrastructure projects, which are concentrated in one part of a city, such as in Berlin's district of Friedrichshain-Kreuzberg, we tend to underestimate the effect. 

Our estimation sample contains 22 treated cities and 84 control cities, both of which some are dropped from our Poisson regressions depending on the specification because of a lack of variation after removing fixed effects or because we do not have observations for our control variables. Fig. S2. shows the 20 treatment cities, for which the size of pop-up infrastructure projects has been recorded in kilometre. We can see that Dublin and Berlin have been the earliest adopters of pop-up bike lanes in the sample and Paris has been the city with the largest program. We use this variation in both timing and the extent of the implemented infrastructure to estimate our effects. We also include control cities in the chart to illustrate the distribution of control cities across European countries. We have a large sample from both France and Germany. This allows us to estimate our effect based on within-country variation removing time-varying factors related to the pandemic that could create bias in our estimates. Note that, while important cities such as London, Milan, Lisbon and Rome had either announced or already implemented a pop-up bike lane program at the time of the analysis, they are missing from the sample due to insufficient spatial or temporal coverage of the bike count data. The average length by city of all bike infrastructures in our sample combined is 11.5 kilometres, the length of bike lanes is 8.2 and the number of measures implemented 19.8. 

We check the sensitivity of our results to different specifications of the treatment, for instance as an indicator variable that is 1, if there is any cycling related infrastructure change in a city and 0 otherwise.
The average effect of having any pop-up infrastructure treatment in a city is 6\% (column 4 in Table S3). The effect per individual measure taken by cities is 0.4\% (column 3). Our findings are robust, when we define treatment based exclusively on those projects that are clearly marked as bike lanes in the data (column 2) rather than based on all types of cycling and traffic calming measures combined (column 1).

\paragraph{Mobility and weather controls}
Our identification strategy relies on the use of different control groups that we expect to be on a common trend around individual daily cohorts of pop-up infrastructure projects. As a baseline we remove and therefore control for time-invariant differences between cities and the locations of the individual counters in our data. Therefore any additional time-invariant control variables would be redundant in our analysis. We also use fixed effect interacting different spatial levels with time dimensions, thereby controlling for many time-varying factors.
We use additional data that varies at a high spatial and temporal resolution to rule out any bias that may be introduced by time-varying factors below our fixed effect levels.

We use weather data from the ERA5 climate model, that provides hourly reanalysis measures of surface temperature, UV radiation, precipitation and wind at a $0.25\degree\times0.25\degree $ resolution\autocite{hersbach2020}. We use the ecwmfr package\autocite{hufkens2019} to aggregate this to the EU Urban Audit city polygons at the daily level. We capture average human mobility throughout the phase of the COVID-19 pandemic starting in March with a human mobility index based on Facebook data\autocite{facebook2020}. The index is from a data set called "movement range maps" that Facebook shares after aggregating individual user movements for humanitarian and research purposes with a reference to the principles outlined by epidemiologists and public health researchers\autocite{buckee2020}. It measures the number of daily 600 meter grid cells visited by Facebook users compared to a baseline in February. For most of our sample the index is aggregated to the state-level, where we use the data. Table S1 shows that on average in our sample period daily mobility has been below the February baseline. 

\paragraph{Regression model}
We model the relationship between cycling traffic and the pop-up bike lane treatment as:

\begin{equation}\centering{
\log \text{Bike Count}_{i c d}=\beta_{1} \text {Bike Lane (km)}_{c d}+\mathbf{X}_{i d}+\lambda_{i} + \sigma_{c w  }+\varphi_{n d}+\varepsilon_{i d}
}
\end{equation}

where ${i}$ indexes a counter, ${c}$ indexes a city, ${n}$ indexes a country, ${d}$ indexes a day, and ${w}$ indexes as week. 

$\lambda_{i}$ is a counter fixed effect that controls for time invariant factors at a high spatial resolution. $\sigma_{c w}$ is a city-week fixed effect that controls for week-specific time-varying factors effectively restricting identifying variation to days before and after treatment within the same week in the same city. $\varphi_{n d}$ is a country-day fixed effect that captures any daily changes common to all cities in a country.

The coefficient of interest is $\beta_{1}$. It captures the effect of the pop-up bike lane treatment on average bicycle counts in a city. Our baseline treatment variable is defined as the number of kilometre of pop-up bike lanes implemented on a given day. Multiplied by a 100 the estimate can be interpreted as the change in bicycle count for a unit change in the treatment variable.

$\mathbf{X}_{i d}$ is a vector of control variables including the mobility index based on Facebook data, weather variables (temperature, UV radiation, wind, precipitation) and the number of counters per city.

We use Poisson pseudo-maximum likelihood regressions (PPML) to estimate this model\autocite{correia2020}. As a robustness check we also use ordinary least squares (OLS) with the natural logarithm of the bicycle count as the outcome (see Fig. 1).
We cluster standard errors at the city-level, where treatment is assigned\autocite{abadie2017}.

\paragraph{Calculating the health benefits of the policy}
We calculate the health benefits by combining our estimates of cycling increases for each kilometre of pop-up bike lane with an estimate of the average health benefits of a kilometre cycled (\$0.45 converted from 0.62 Australian Dollars)\autocite{zapata-diomedi2018}, which is lower than typical values from the grey literature\autocite{zapata-diomedi2018}. Our regression estimates only provide us with a percentage increase in cycling (0.6\%) per kilometre of bike lane. We convert this result into additional kilometres cycled in a city based on baseline values of kilometres cycled per person in a city from a detailed transport behaviour survey in 135 German cities\autocite{hubrich2019}. We impute values of kilometres cycled for other European cities based on information on the modal split (trips) of commutes \autocite{eurostat2020b} and a city's population both taken from the European Urban Audit\autocite{eurostat2020c}.

Our estimate only counts benefits from cycling but not the saved costs of a potential modal shift from car use to cycling, that we cannot measure with our data. It also does not take into account shifts from walking or jogging and cycling for exercise to cycling in the city, where counters in our sample are typically placed. Since the external costs of car use are high\autocite{gossling2019}, we interpret our calculation as a lower bound.

% "If cycling levels in urban England and Wales increased to levels seen in Denmark, the associated healthcare cost savings have been estimated at £17 billion over twenty years" (Jarrett et al., 2012).

% \paragraph{Comparison to effects from other bike infrastructure interventions}
% \autocite{goodman2013,panter2016,aldred2019b}
%\autocite{mueller2018}

% External validity Selection in our cities, selection where counters are, bike lanes get places on existing counter routes

\clearpage
\begin{table}[h!]
\scriptsize
\centering
\begin{tabular}{l*{1}{cccccccc}}
\toprule
                    &        Mean&  Std.\ Dev.&        25\%&        50\%&        75\%&        95\%&        Min.&        Max.\\
\midrule
Daily number of cyclists&      1457.2&      1895.7&         255&         744&        1923&        5151&           1&       13339\\
City size (ha)      &     32893.8&     42393.6&     14163.3&     22018.5&     40659.9&     89180.2&       455.6&      251517\\
Year                &        2017&           2&        2016&        2018&        2019&        2020&        2007&        2020\\
Number of counters in the same city&        22.9&        23.2&           4&          14&          32&          82&           1&          90\\
Facebook mobility index&       -0.16&        0.21&       -0.27&       -0.11&     -0.0044&       0.088&       -0.81&        0.51\\
\midrule
Observations        &      995818&            &            &            &            &            &            &            \\
\bottomrule
\end{tabular}
\caption*{\scriptsize{\textbf{Table S1. Summary statistics at the counter-day level} The unit of observation of our analysis is the counter and data varies daily. Count data is from municipal bike counters and is obtained from different APIs. Treatment and control variables are assigned to counters based on their city attribute. City definitions are from the EU Urban Audit. The Facebook mobility index is only available from March 2020. It measures aggregate movement activity by Facebook users in a given administrative area (districts or states).}} 
\end{table}

\clearpage

\begin{table}[h!]
\scriptsize
\centering
\begin{tabular}{l*{1}{ccccccc}}
\toprule
                    &        Mean&  Std.\ Dev.&        25\%&        50\%&        75\%&        95\%&        Max.\\
\midrule
Total length of bike infrastructures&        11.5&        20.0&        1.39&        2.57&        16.6&        57.9&        85.1\\
Total lenght of bike lanes&        8.24&        18.3&        0.24&        2.05&        7.35&        24.8&        84.3\\
Number of measures  &        19.8&        48.1&           1&           4&          17&          52&         226\\
\midrule
Observations        &          22&            &            &            &            &            &            \\
\bottomrule
\end{tabular}
\caption*{\scriptsize{\textbf{Table S2. Summary statistics of most recent state of infrastructure at the city level show} We use data from the European Cyclists' Federation. The raw data includes information on individual infrastructure projects announced or implemented. We aggregate it to the city-day level using city definitions from the EU Urban Audit. Our analysis includes data up to the 8 July 2020. The newest data can be found at: \url{https://datastudio.google.com/u/0/reporting/ba90a08c-9841-4beb-9e26-7d4f7d002709/page/yMRTB}}} 
\end{table}

\clearpage

\begin{table}[h!]
\scriptsize
    \centering
\begin{tabular}{l*{4}{S}}
\toprule
          &\multicolumn{4}{c}{Outcome: Cyclist count}                                 \\\cmidrule(lr){2-5}
          &\multicolumn{1}{c}{(1)}&\multicolumn{1}{c}{(2)}&\multicolumn{1}{c}{(3)}&\multicolumn{1}{c}{(4)}\\
          &\multicolumn{1}{c}{\shortstack{All km}}&\multicolumn{1}{c}{\shortstack{Bike lane km}}&\multicolumn{1}{c}{\shortstack{Num of measures}}&\multicolumn{1}{c}{\shortstack{Any treatment}}\\
\midrule
Pop-up treatment&    0.006\sym{**} &    0.007\sym{**} &    0.004\sym{*}  &    0.061\sym{*}  \\
          &  (0.003)         &  (0.003)         &  (0.002)         &  (0.036)         \\
\midrule
City clusters&\multicolumn{1}{c}{78}         &\multicolumn{1}{c}{78}         &\multicolumn{1}{c}{78}         &\multicolumn{1}{c}{78}         \\
N         &\multicolumn{1}{c}{59904}         &\multicolumn{1}{c}{59904}         &\multicolumn{1}{c}{59904}         &\multicolumn{1}{c}{59904}         \\
\bottomrule
\end{tabular}
\caption*{\scriptsize{\textbf{Table S3. Different treatment specifications} Each column shows the effect of treatment with pop-up infrastructure on a city's cycling count compiled from city APIs. The data on daily pop-up bike lane additions are from the European Cyclists' Federation\autocite{ecf2020}. The unit of observation is the cycling counter. Time variation is daily. Coefficients are from Poisson regressions. Column (1) shows the effect of a kilometre of any bike infrastructure, (2) shows the effect of bike lanes, (3) the effect of any single measure in a city, and (4) the overall treatment of an implemented pop-up infrastructure program in a city.
All regressions include counter and day fixed effects and controls for overall mobility (measured with Facebook user movements), weather (temperature, wind, sunshine, precipitation), and the number of counters in a city. We cluster standard errors at the city level, where treatment is assigned. Significance levels are \sym{*} p $<$ 0.1, \sym{**} p $<$ 0.05, \sym{***} p $<$ 0.01.}}
\end{table}

\clearpage
\begin{figure}[h!]
\centering
    \includegraphics[width=\textwidth]{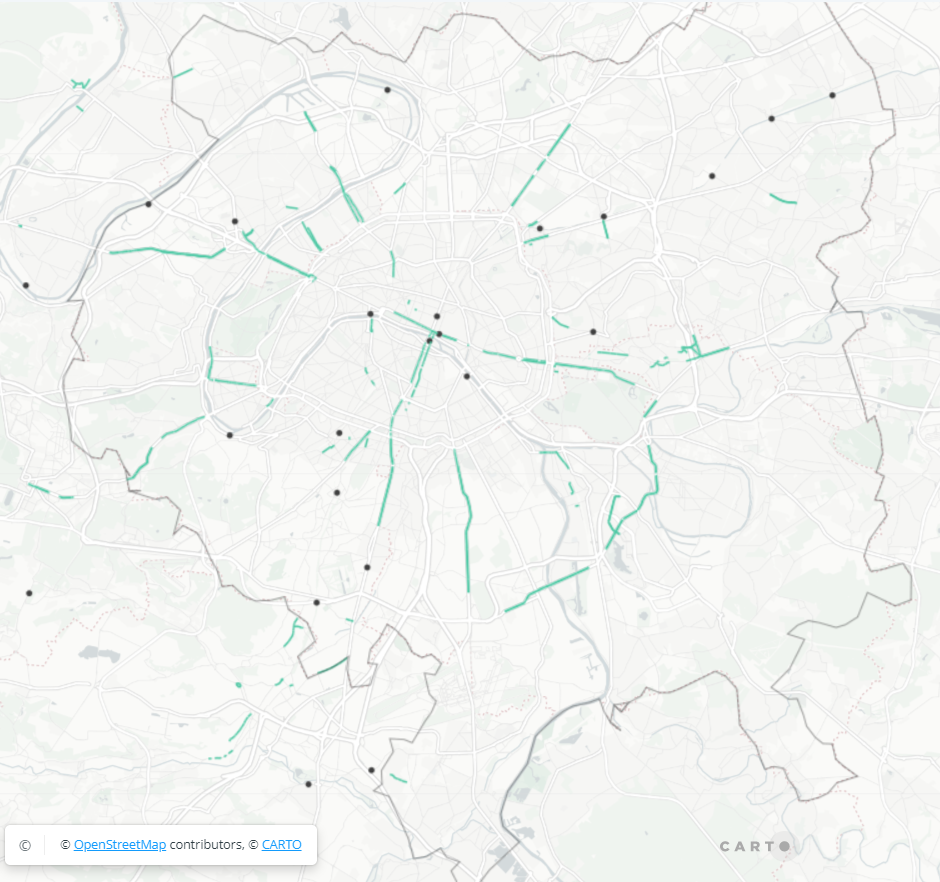}
    \caption{\scriptsize{{\bf Fig. S1. Pop-up bike lanes and bicycle counters in Paris} The map shows pop-up bike lanes implemented in Paris up to 3 July 2020 (green lines) and the location of bike counters (dots) in our data set. The detailed infrastructure data has been collected by a consortium of French NGOs and researchers. It is available at: \url{https://carto.parlons-velo.fr/\#10.13/48.8312/2.5506}}}
\end{figure}

\begin{figure}[h!]
\centering
    \includegraphics[width=0.5\textwidth]{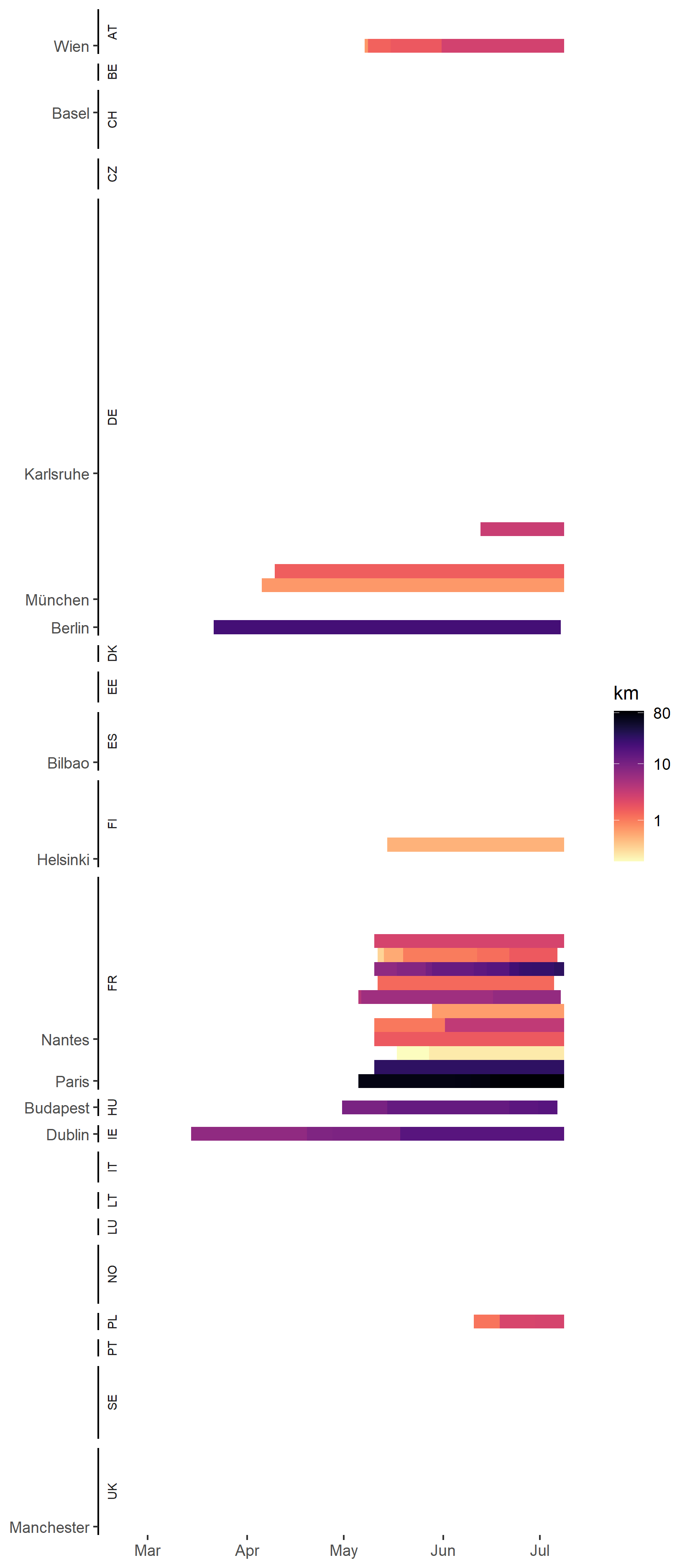}
    \caption{\scriptsize{{\bf Fig. S2. Intensity of pop-up bike lane treatment over time in treatment cities and control cities} This Figure shows treated cities and their treatment intensities in implemented kilometres (colouring on a log scale) of public bike lanes at a given day between March and July 2020. Control cities are plotted in white. London, Milan, Lisbon and Rome are missing from the sample due to insufficient spatial or temporal coverage of the data.
    Information on individual pop-up bike lanes with their street location, announcement  implementation is from the European Cyclists' Federation. The newest data can be found at: \url{https://ecf.com/dashboard}}}
\end{figure}

% \begin{landscape}
% \begin{figure}[h!]
% \centering
%     \includegraphics[width=1.3\textwidth]{figuresSI/publSpecsLarge.pdf}
%     \caption{\scriptsize{{\bf Fig. S3. Effect of pop-up bike lanes on cycling in different model specifications} This figure shows estimates of regressions of the daily cyclist count on the number of kilometres of pop-up bike lane implemented at a given day in a city. The unit of observation is the bike counter. Baseline specifications are marked in blue. Darker colors in the bottom panel indicate the type of specification used. The 95\% confidence interval is shown in darker colour and the 90\% confidence interval in lighter colour. The estimates can be interpreted as the average increase in the level of cycling caused by one kilometre of pop-up bike lane. Control variables are from Facebook (mobility index measured by user movements) and the ERA5 climate model (weather variables). The variable "Number of counters" indicates the total of counters per city.
%     }}
% \end{figure}
% \end{landscape}

\begin{figure}[h!]
\centering
    \includegraphics[width=\textwidth]{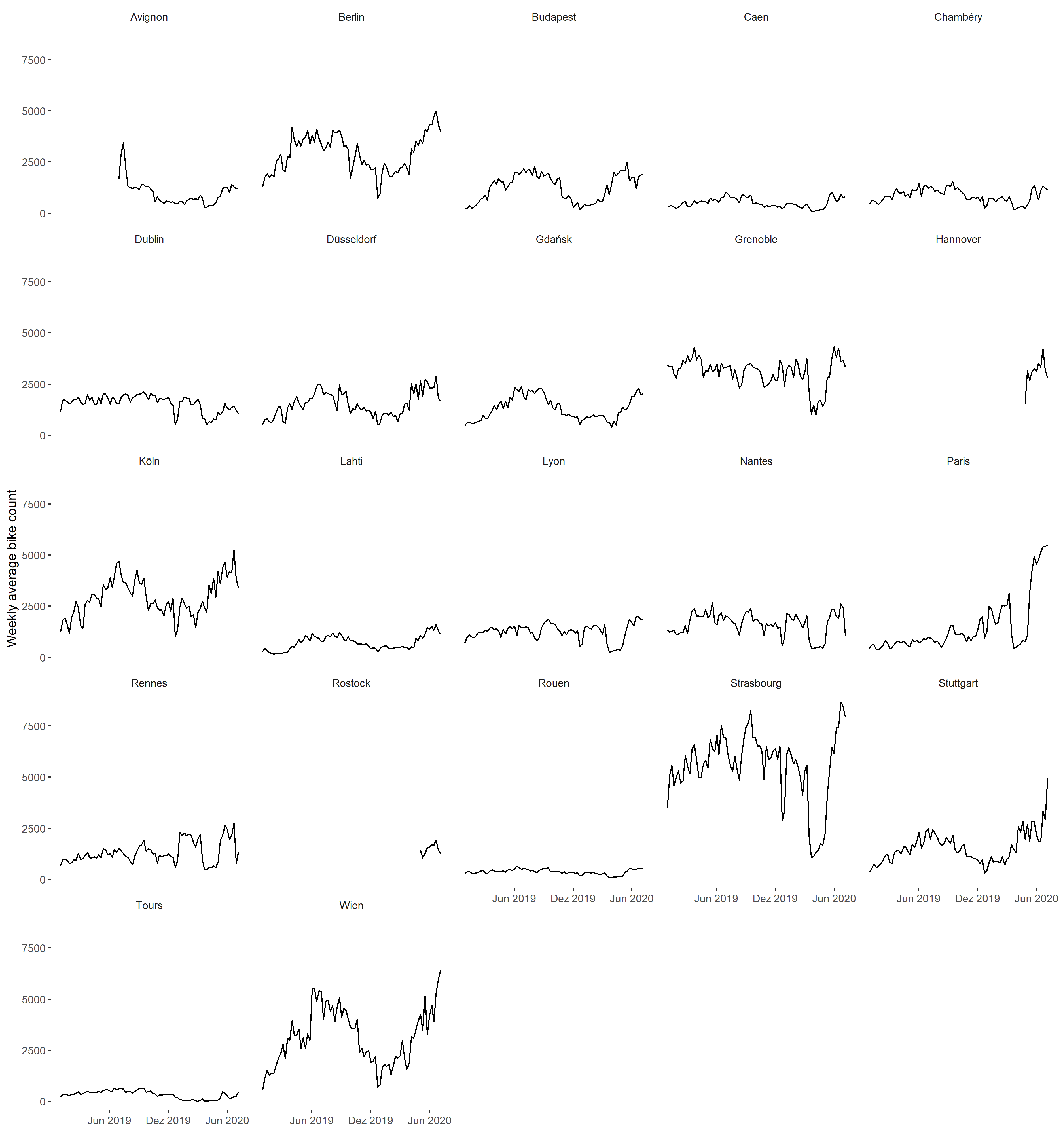}
    \caption{\scriptsize{{\bf Fig. S3. Average bike count per week in treated cities} Daily bike counts are aggregated by city and averaged over the week. Bike counts are assembled from municipal open data feeds. The lower and upper percentiles from the initial sample (treated and control cities combined) are removed from the sample. Only measurements from 2019 and 2020 are shown. City definitions are chosen according to EU Urban Audit.
    }}
\end{figure}

\begin{figure}[h!]
\centering
    \includegraphics[width=\textwidth]{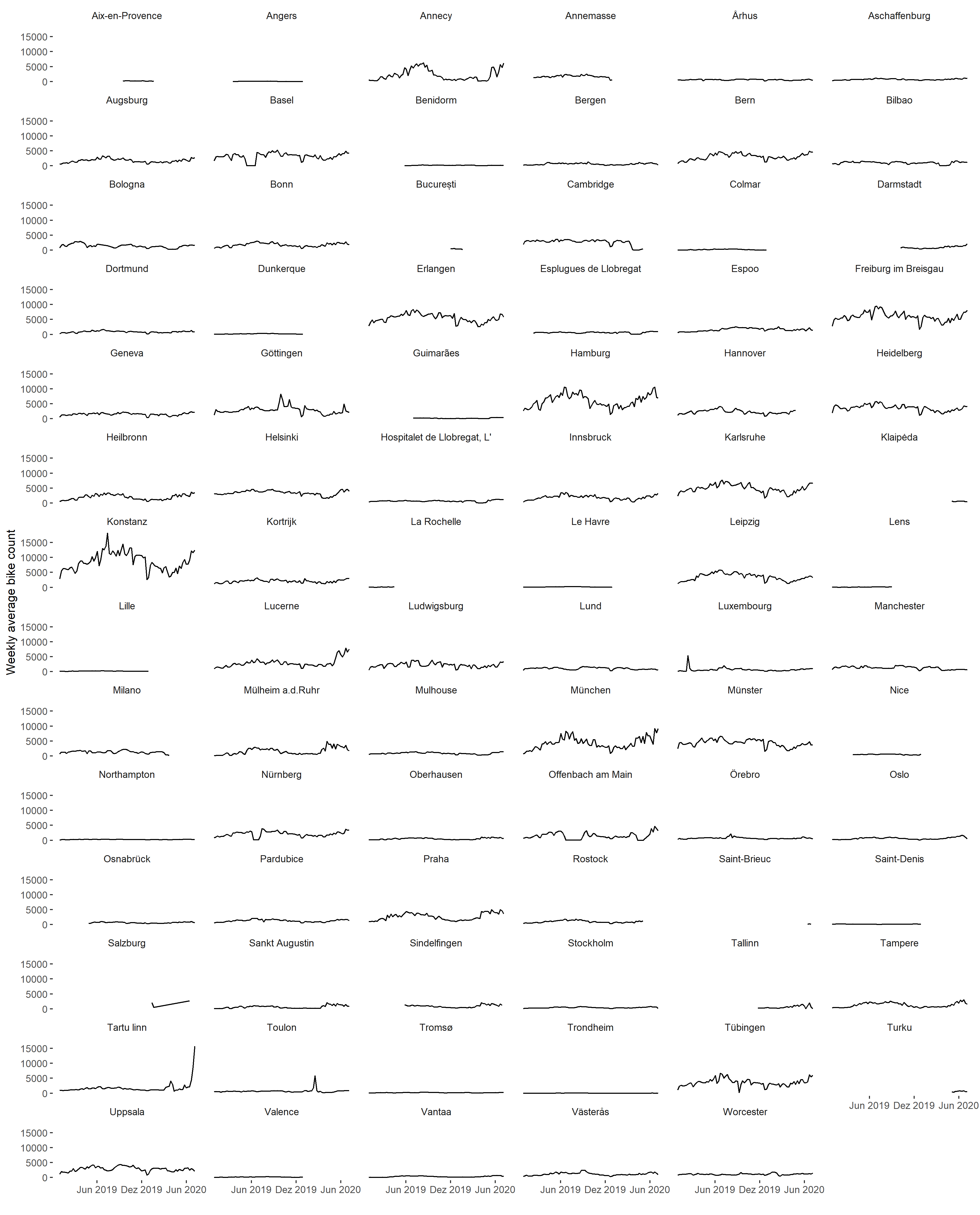}
    \caption{\scriptsize{{\bf Fig. S4. Average bike count per week in control cities} Daily bike counts are aggregated by city and averaged over the week. Bike counts are assembled from municipal open data feeds. The lower and upper percentiles from the initial sample (treated and control cities combined) are removed from the sample. Only measurements from 2019 and 2020 are shown. City definitions are chosen according to EU Urban Audit.
    }}
\end{figure}

\clearpage
\newpage
\printbibliography
\end{refsection}
\end{document}